\documentclass[prd,aps,showpacs,nofootinbib]{revtex4}%
\usepackage{amssymb}
\usepackage{graphicx}
\usepackage{amsmath}
\usepackage{amsfonts}%
\newcommand{\be}{\begin{equation}}
\newcommand{\ee}{\end{equation}}
\newcommand{\bq}{\begin{eqnarray}}
\newcommand{\eq}{\end{eqnarray}}
\begin{document}
\title{Electron-electron interaction in a MCS model with a purely spacelike
Lorentz-violating background}
\author{M.M. Ferreira Jr.}
\affiliation{{\small {Universidade Federal do Maranh\~{a}o (UFMA), Departamento de
F\'{\i}sica, Campus Universit\'{a}rio do Bacanga, S\~{a}o Luiz - MA, 65085-580
- Brasil}}}

\begin{abstract}
One considers a planar Maxwell-Chern-Simons electrodynamics in the presence of
a purely spacelike Lorentz-violating background. Once the Dirac sector is
properly introduced and coupled to the scalar and the gauge fields, the
electron-electron interaction\ is evaluated as the Fourier transform of the
M\"{o}ller scattering amplitude (derived in the non-relativistic limit). The
associated Fourier integrations can not be exactly carried out, but an
algebraic solution for the interaction potential is obtained in leading order
in v$^{2}/s^{2}.$ It is then observed that the scalar potential presents a
logarithmic attractive (repulsive) behavior near (far from) the origin.
Concerning the gauge potential, it is composed of the pure MCS interaction
corrected by background contributions, also responsible for its anisotropic
character.\ It is also verified that such corrections may turn the gauge
potential attractive for some parameter values. Such attractiveness remains
even in the presence of the centrifugal barrier and gauge invariant $A\cdot A$
term, which constitutes a condition compatible with the formation of Cooper pairs.

\end{abstract}
\maketitle

\section{Introduction}

In the latest years, Lorentz-violating theories have been in focus of great
interest and investigation \cite{Kostelec1}-\cite{Manojr2}. Despite the
intensive activity proposing and discussing the consequences of a
Lorentz-violating electrodynamics, some experimental data and theoretical
considerations indicate stringent limits on the parameters responsible for
such a breaking \cite{Coleman}, \cite{Particles}. These evidences put the
Lorentz-violation as a negligible effect in a factual (1+3)-dimensional
electrodynamics, which raises the question about the feasibility of
observation of this effect in a lower dimension system or in another
environment distinct from the usual high-energy domain in which this matter
has been generally regarded so far.

Condensed Matter Systems (CMS)\ are low-energy systems sometimes endowed with
spatial anisotropy which might constitute a nice environment to study
Lorentz-violation and to observe correlated effects. Indeed, although Lorentz
covariance is not defined in a CMS, Galileo covariance holds as a genuine
symmetry in such a system (a least for the case of isotropic low-energy
systems). Having in mind that a CMS may be addressed as the low-energy limit
of a relativistic model, there follows a straightforward correspondence
between the breakdown of Lorentz and Galileo symmetries, in the sense that a
CMS with\ violation of Galileo symmetry may have as counterpart a relativistic
system endowed with breaking of Lorentz covariance.\thinspace\ Considering the
validity of this correspondence, it turns out that an anisotropic CMS may be
addressed as the low-energy limit of a relativistic model in the presence of a
spacelike Lorentz-violating background.

\ The attainment of an attractive electron-electron $(e^{-}e^{-})$\ potential
in the context of a planar model incorporating Lorentz-violation is a point
that sets up a clear connection between such theoretical models and condensed
matter physics. Theoretical planar models able to provide attractive
$e^{-}e^{-}$\ interaction\ potentials may constitute a suitable framework to
deal with the condensation of Cooper pairs, a fundamental characteristic of
superconducting systems. The Maxwell-Chern-Simons theories \cite{MCS} were
addressed in the beginning 1990s as a theoretical alternative to accomplish
this objective, without success. In a recent calculation \cite{Ghosh}, new
results concerning an electron-electron interaction were also obtained in the
context of a noncommutative extension of the MCS electrodynamics, revealing a
non-relativistic potential nearly identical to the MCS outcome. Actually, it
is known that the MCS-Proca models \cite{Tese} may better provide an
attractive interaction due to the action of the intermediation played by the
Higgs sector. Another well defined feature of a planar superconductor concerns
the symmetry of the order parameter (standing for the Cooper pair), which is
described in terms of a spatially anisotropic d-wave \cite{OP}. A field theory
model able to account for an anisotropic $e^{-}e^{-}$\ interaction is the
first step to the achievement of anisotropy for the order parameter. This is
exactly the expected result to be obtained in the case of a Lorentz-violating
model in the presence of a purely spacelike background, where the $e^{-}e^{-}%
$\ scattering potential may be identified with the one evaluated in the
context of a CMS endowed with a privileged direction in space.

The investigation of the $e^{-}e^{-}$\ interaction can be suitably considered
in the context of a Lorentz-violating planar framework. \ In fact, in a very
recent paper \cite{Manojr3}, the low-energy M\"{o}ller interaction potential
was carried out for the case of a planar electrodynamics \cite{Manojr1}
arising from the dimensional reduction of the Carroll-Field-Jackiw model
\cite{Jackiw}, for a purely timelike background. With this purpose, the Dirac
sector was included in this planar model, so that to make feasible the
consideration of the low-energy M\"{o}ller scattering (adopted as the
appropriate tool to analyze the non-relativistic electron-electron
interaction). The interaction potential obtained revealed to be composed of a
scalar and a gauge contributions. The scalar one (coming from the scalar
intermediation) has presented a logarithmic attractive (repulsive) behavior
near (far from) the origin. On the other hand, it has been shown that gauge
potential, associated with the gauge intermediation, is composed of the
Maxwell-Chern-Simons (MCS)\ usual interaction \cite{MCS} corrected by
background depending terms. One has also noted that these corrections lead to
a gauge potential endowed with attractiveness (for some parameters values)
even in the presence of the centrifugal barrier and the $\mathbf{A}^{2}-$gauge
invariant term stemming from the Pauli equation. Thereby, one has shown that
these results bypass the controversy involving the pure Maxwell-Chern-Simons
potential (see Hagen and Dobroliubov \cite{MCS}) concerning the possibility of
attractiveness, and yields a strong indication that it may occur the formation
of Cooper pairs in this theoretical framework.

Having as main motivation the encouraging outcomes achieved in Ref.
\cite{Manojr3}, in this work one searches for the electron-electron potential
in the context of a Lorentz-violating planar electrodynamics endowed with a
purely spacelike background, $v^{\mu}=(0,\mathbf{v)}$. As this kind of
background fixes a 2-direction in space, it will certainly lead to an
anisotropic behavior, one consequence of the directional dependence of the
solutions in relation to the fixed background $\left(  \mathbf{v}\right)  $.
By determining such $e^{-}e^{-}$ potential, one can investigate two expected
properties concerning the $e^{-}e^{-}$ interaction: attractiveness and
anisotropy, which are relevant due its possible connection with high-T$_{c}$
superconducting systems. \ The procedure here adopted is the same one
developed in Ref. \cite{Manojr3}, that is, one carries out the $e^{-}e^{-}%
$\ interaction potential stemming from the M\"{o}ller scattering amplitude
associated with the scalar and the gauge intermediations, exhibiting and
pointing out the corrections induced by the fixed background. on the pure
Maxwell-Chern-Simons result. With this purpose, one starts from the planar
Lagrangian defined in Ref. \cite{Manojr3}, in which the Dirac sector has been
already included. One then carries out the $e^{-}e^{-}$ M\"{o}ller scattering
amplitude (from which the interaction potential is derived according to the
Born approximation) following the general guidelines set up in Refs.
\cite{MCS}, \cite{Tese}, \cite{Manojr3}. The potential here attained is
composed of a scalar and a gauge contribution as well, since the $e^{-}e^{-}%
$\ interaction is equally mediated by the massless scalar and the massive
gauge fields. The scalar potential maintains the logarithmic behavior
(asymptotically repulsive and attractive near the origin) of the purely
timelike case, being different only by the presence of anisotropy. As for the
gauge potential, it is given by a lengthy expression composed of the pure MCS
interaction and many background-depending terms which imply the presence
anisotropy, among other features. It is possible to show that these
corrections are able to turn this potential attractive for some values of the
relevant parameters, behavior which remains even in the presence of the
centrifugal barrier $\left(  l/mr^{2}\right)  $ and the $\mathbf{A}%
\cdot\mathbf{A}$ gauge invariant term. Furthermore, the total interaction
(scalar and gauge potentials) may always be attractive with a suitable adjust
of the coupling parameters. This outcome constitute the essential condition to
promote the condensation of Cooper pairs, which shows that this theoretical
framework may be useful to address some properties of superconducting systems.

This paper is organized as follows. In Sec. II, one briefly presents the
structure of reduced planar model (derived in Ref. \cite{Manojr1}) which is
here adopted as stating point. This model is supplemented by the Dirac field.
In Sec. III, one presents the spinors which fulfill the two-dimensional Dirac
equation and that are used to evaluate the M\"{o}ller scattering amplitude
associated with the Yukawa and the gauge intermediations. The corresponding
interaction potentials are carried out, and the results are discussed. In Sec.
IV, one concludes with the Final Remarks.

\section{The planar Lorentz-violating model}

The starting point is the planar Lagrangian obtained from the dimensional
reduction of the CFJ-Maxwell electrodynamics \cite{Manojr1}, which consists in
a Maxwell-Chern-Simons electrodynamics coupled to a massless scalar field
$\left(  \varphi\right)  $ and to a fixed background 3-vector $\left(  v^{\mu
}\right)  $ through a Lorentz-violating term, derived from the dimensional
reduction of the Carroll-Field-Jackiw model \cite{Jackiw}. One then regards
the additional presence of a fermion field $\left(  \psi\right)  $:%

\begin{align}
\mathcal{L}_{1+2}  &  =-\frac{1}{4}F_{\mu\nu}F^{\mu\nu}+\frac{s}{2}%
\epsilon_{\mu\nu\kappa}A^{\mu}\partial^{\nu}A^{\kappa}-\frac{1}{2}%
\partial_{\mu}\varphi\partial^{\mu}\varphi+\varphi\epsilon_{\mu\nu\kappa
}v^{\mu}\partial^{\nu}A^{\kappa}-\frac{1}{2\alpha}\left(  \partial_{\mu}%
A^{\mu}\right)  ^{2}\nonumber\\
&  +{\overline{\psi}(i}{\rlap{\hbox{$\mskip 4.5 mu /$}}D}-m_{e}){\psi
}-y\varphi(\overline{\psi}\psi). \label{L1}%
\end{align}
Here, the covariant derivative, ${\rlap{\hbox{$\mskip 4.5
mu /$}}D}\psi\equiv(\rlap{\hbox{$\mskip 1
mu /$}}\partial+ie_{3}\rlap{\hbox{$\mskip 3 mu /$}}A)\psi,$ states the minimal
coupling, whereas the term $\varphi(\overline{\psi}\psi)$ reflects the Yukawa
coupling between the scalar and fermion fields, with $y$ being the constant
that measures the strength of the electron-phonon coupling. The mass
dimensions of the fields and parameters are the following: $\left[
\varphi\right]  =\left[  A^{\mu}\right]  =1/2,\left[  \psi\right]  =1,\left[
s\right]  =\left[  v^{\mu}\right]  =1,\left[  e_{3}\right]  =\left[  y\right]
=1/2.$ One hen notes that the coupling constants, $e_{3}$, $y,$ both exhibit
$\left[  mass\right]  ^{1/2}$ dimension, a usual result in (1+2) dimensions.
Furthermore, in Ref. \cite{Manojr1} the propagators of the scalar $\left(
\varphi\right)  $ and gauge $\left(  A_{\mu}\right)  $ fields were properly
evaluated and used as starting point to analyze the causality and unitarity.
Such analysis has revealed a model totally stable, causal and unitary (at a
classical level) for both spacelike and timelike backgrounds. This result has
demonstrated that this planar model bypasses the problems concerning the
stability and causality presented by the original Carroll-Field-Jackiw (CFJ)
model \cite{Jackiw} in (1+3) dimensions, which indicates that this model may
undergo consistent quantization procedures, a necessary condition to address
condensed matter systems. The knowledge of the propagators\footnote{The gauge
propagator is given by: $\langle A^{\mu}\left(  k\right)  A^{\nu}\left(
k\right)  \rangle=i\biggl\{-\frac{1}{k^{2}-s^{2}}\theta^{\mu\nu}-\frac
{\alpha(k^{2}-s^{2})\boxtimes(k)+s^{2}\left(  v.k\right)  ^{2}}{k^{2}%
(k^{2}-s^{2})\boxtimes(k)}\omega^{\mu\nu}-\frac{s}{k^{2}(k^{2}-s^{2})}%
S^{\mu\nu}+\frac{s^{2}}{(k^{2}-s^{2})\boxtimes(k)}\Lambda^{\mu\nu}-\frac
{1}{(k^{2}-s^{2})\boxtimes(k)}T^{\mu}T^{\nu}+\frac{s}{(k^{2}-s^{2}%
)\boxtimes(k)}[Q^{\mu\nu}-Q^{\nu\mu}]+\frac{is^{2}\left(  v.k\right)  }%
{k^{2}(k^{2}-s^{2})\boxtimes(k)}[\Sigma^{\mu\nu}+\Sigma^{\nu\mu}%
]-\frac{is\left(  v.k\right)  }{k^{2}(k^{2}-s^{2})\boxtimes(k)}[\Phi^{\mu\nu
}-\Phi^{\nu\mu}]\biggr\},$ whereas the scalar propagator is: $\langle
\varphi\varphi\rangle=\frac{i}{\boxtimes(k)}\left[  k^{2}-s^{2}\right]  ,$
where: $\boxtimes(k)=\left[  k^{4}-\left(  s^{2}-v\cdot v\right)
k^{2}-\left(  v\cdot k\right)  ^{2}\right]  $. The involved 2-rank tensors are
defined as follows: $\theta_{\mu\nu}=\eta_{\mu\nu}-\omega_{\mu\nu},$%
\ $\omega_{\mu\nu}=\partial_{\mu}\partial_{\nu}/\square,$ $S_{\mu\nu
}=\varepsilon_{\mu\kappa\nu}\partial^{\kappa},$ $Q_{\mu\nu}=v_{\mu}T_{\nu
},T_{\nu}=S_{\mu\nu}v^{\mu},$ $\Lambda_{\mu\nu}=v_{\mu}v_{\nu},$
\ $\Sigma_{\mu\nu}=v_{\mu}\partial_{\nu},$ $\Phi_{\mu\nu}=T_{\mu}\partial
_{\nu}.$} evaluated in Ref. \cite{Manojr1} is essential to the calculations of
this work.

\section{The M\"{o}ller Scattering amplitude and the interaction potential}

The two-particle interaction potential is given by the Fourier transform of
the two-particle scattering amplitude in the low-energy limit (Born
approximation). In the case of the nonrelativistic M\"{o}ller scattering, one
should consider only the t-channel (direct scattering) \cite{Sakurai} even for
indistinguishable electrons, since in this limit one recovers the classical
notion of trajectory. From Eq. (\ref{L1}), there follow the Feynman rules for
the interaction vertices: $V_{\psi\varphi\psi}=iy;V_{\psi A\psi}=ie_{3}%
\gamma^{\mu}$, so that the $e^{-}e^{-}$ scattering amplitude are written as:
\begin{align}
-i\mathcal{M}_{scalar}  &  =\overline{u}(p_{1}^{^{\prime}})(iy)u(p_{1})\left[
\langle\varphi\varphi\rangle\right]  \overline{u}(p_{2}^{^{\prime}%
})(iy)u(p_{2}),\label{A1}\\
-i\mathcal{M}_{A}  &  =\overline{u}(p_{1}^{^{\prime}})(ie_{3}\gamma^{\mu
})u(p_{1})\left[  \langle A_{\mu}A_{\nu}\rangle\right]  \overline{u}%
(p_{2}^{^{\prime}})(ie_{3}\gamma^{\nu})u(p_{2}), \label{A2}%
\end{align}
with $\langle\varphi\varphi\rangle$ and $\langle A_{\mu}A_{\nu}\rangle$ being
the scalar and photon propagators. Expressions (\ref{A1}) and (\ref{A2})
represent the scattering amplitudes for electrons of equal polarization
mediated by the scalar and gauge particles, respectively. The spinors $u(p)$
stand for the positive-energy solution of the Dirac equation $\left(
\rlap{\hbox{$\mskip1 mu /$}}p-m\right)  u(p)=0$. The $\gamma-$ matrices
satisfy the $so(1,2)$ algebra, $\left[  \gamma^{\mu},\gamma^{\nu}\right]
=2i\epsilon^{\mu\nu\alpha}\gamma_{\alpha}$, and correspond to the
(1+2)-dimensional representation of the Dirac matrices, that is, the Pauli
ones: $\gamma^{\mu}=(\sigma_{z},-i\sigma_{x},i\sigma_{y}).$ Regarding these
definitions, one obtains the spinors,%

\begin{equation}
u(p)=\frac{1}{\sqrt{N}}\left[
\begin{array}
[c]{c}%
E+m\\
-ip_{x}-p_{y}%
\end{array}
\right]  ,\text{ \ \ }\overline{u}(p)=\frac{1}{\sqrt{N}}\left[
\begin{array}
[c]{cc}%
E+m & -ip_{x}+p_{y}%
\end{array}
\right]  , \label{spinor}%
\end{equation}
which fulfill the normalization condition $\overline{u}(p)u(p)=1$ whenever the
constant $N=2m(E+m)$ is adopted. The M\"{o}ller scattering should be easily
analyzed in the center of mass frame, where the momenta of the incoming and
outgoing electrons are read at the form: \ $P_{1}^{\mu}=(E,p,0),P_{2}^{\mu
}=(E,-p,0),P_{1}^{^{\prime}\mu}=(E,p\cos\theta,p\sin\theta),P_{2}^{^{\prime
}\mu}=(E,-p\cos\theta,-p\sin\theta),$whereas $\theta$ is the scattering angle
(in the CM frame)$.$ The 3-current components, $j^{\mu}(p)=\overline
{u}(p^{^{\prime}})\gamma^{\mu}u(p),$\ and the transfer 3-momentum arising from
this convention are explicitly written in Ref. \cite{Manojr3}.

\subsection{The scalar potential}

Starting from the expression of the scalar propagator $\langle\varphi
\varphi\rangle$ (see footnote 1), considering the transfer momentum, $k^{\mu
}=(0,\mathbf{k}),$ and a purely spacelike background, $v^{\mu}=(0,\mathbf{v})
$, the following scattering amplitude arises from Eq. (\ref{A1}):
\begin{equation}
\mathcal{M}_{scalar}=-y^{2}\frac{\left[  \mathbf{k}^{2}+s^{2}\right]
}{\mathbf{k}^{2}\left[  \mathbf{k}^{2}+s^{2}+\text{v}^{2}\sin^{2}%
\alpha\right]  }, \label{Mscalar3}%
\end{equation}
where $\alpha$ is the angle defined by the vectors $\mathbf{v}$ and
$\mathbf{k}$. Taking into account the Born approximation, the potential
associated with the Yukawa interaction reads as,
\begin{equation}
V_{scalar}(r)=-\frac{y^{2}}{\left(  2\pi\right)  ^{2}}\int e^{i\overrightarrow
{k}.\overrightarrow{r}}\frac{\left[  \mathbf{k}^{2}+s^{2}\right]  }%
{\mathbf{k}^{2}\left[  \mathbf{k}^{2}+s^{2}+\text{v}^{2}\sin^{2}\alpha\right]
}d^{2}\overrightarrow{k}. \label{V_scalar1}%
\end{equation}
Such Fourier integration above can not be exactly carried out. However, this
integration may be solved in the regime in which $s^{2}>>$v$^{2}.$ As far as
this condition holds, the following approximation,%

\begin{equation}
\frac{\left[  \mathbf{k}^{2}+s^{2}\right]  }{\mathbf{k}^{2}\left[
\mathbf{k}^{2}+s^{2}+\text{v}^{2}\sin^{2}\alpha\right]  }\simeq\frac
{1}{\mathbf{k}^{2}}-\frac{\text{v}^{2}\sin^{2}\alpha}{\mathbf{k}^{2}\left[
\mathbf{k}^{2}+s^{2}\right]  }, \label{V_scalar2}%
\end{equation}
is valid in first order in v$^{2}/s^{2}.$ In order to solve Eq.
(\ref{V_scalar1}), two other angles are of interest: $\varphi$ and $\beta$ -
defined respectively by the relations: $\cos\varphi=\mathbf{r}\cdot
\mathbf{k}/rk,$ $\cos\beta=\mathbf{r\cdot v}/r$v$.$ While the background
vector,\ $\mathbf{v},$ sets up a fixed direction in space, the coordinate
vector, $\mathbf{r},$ defines the position where the potentials are to be
measured; so, $\beta$ is the (fixed) the angle that indicates the directional
dependence of the fields in relation to the background direction. Being
confined into the plane, these angles satisfy a simple relation:
$\alpha=\varphi-\beta,$ whose consideration leads to:\ $\sin^{2}\alpha
=c_{2}+c_{1}\cos^{2}\varphi+c_{3}\sin2\varphi,$ with: $c_{1}=(1-2\cos^{2}%
\beta),$ $c_{2}=\cos^{2}\beta,$ $c_{3}=-(\sin2\beta)/2.$ This expression
allows the evaluation of the angular integration on the $\varphi-$variable
enclosed in Eq. (\ref{V_scalar1}), given below:
\begin{equation}
\int_{0}^{2\pi}e^{ikr\cos\varphi}\sin^{2}\alpha d\varphi=2\pi\left[  \left(
c_{1}+c_{2}\right)  J_{0}(kr)-\frac{c_{1}}{kr}J_{1}(kr)\right]  .
\label{Angular1}%
\end{equation}
Taking into account these preliminary results, one shall now carry out the
integrations on the $\mathbf{k}$-variable, obtaining the following scalar
interaction potential:%

\begin{equation}
V_{scalar}(r)=\frac{y^{2}}{\left(  2\pi\right)  }\biggl\{\left[
1-\frac{\text{v}^{2}}{s^{2}}\right]  \ln r-\frac{\text{v}^{2}}{s^{2}}(\sin
^{2}\beta)K_{0}(sr)-\frac{\text{v}^{2}\cos2\beta}{s^{4}}\frac{1}{r^{2}%
}[1-srK_{1}(sr)]\biggr\}. \label{Vscalar2}%
\end{equation}
Near the origin, $r\rightarrow0$, the modified Bessel functions behave as
$K_{0}(r)\rightarrow-\ln r,$ $K_{1}(sr)\rightarrow1/sr+sr\ln r/2,$ apart from
constant terms. In such a way, the potential $V_{scalar}$ goes like:
\begin{equation}
\lim_{r\rightarrow0}V_{scalar}(r)=\frac{y^{2}}{\left(  2\pi\right)  }\left[
1-\frac{\text{v}^{2}}{2s^{2}}(1+\sin^{2}\beta)\right]  \ln r.
\end{equation}
Far from the origin, $r\rightarrow\infty,$ the Bessel functions decay
exponentially whereas the logarithmic function increases. In this limit, one
has:\
\begin{equation}
\lim_{r\rightarrow\infty}V_{scalar}(r)=\frac{y^{2}}{\left(  2\pi\right)
}\left[  1-\frac{\text{v}^{2}}{2s^{2}}\right]  \ln r. \label{Asym}%
\end{equation}
Remaking the condition $\left(  s^{2}>>\text{v}^{2}\right)  $ under which this
solution was derived, the scalar potential turns out always attractive near
the origin and repulsive asymptotically. Thereby, both near and far form the
origin the scalar potential exhibits a logarithmic behavior corrected by the
background term, with explicit directional dependence in terms of the angle
$\beta.$ Such a result implies an attractive (repulsive) interaction near (far
from) the origin. This logarithmic asymptotic behavior also indicates the
absence of screening concerning the scalar sector of this model. The existence
of unscreened solutions is ascribed to the presence of a massless-like term,
$1/[k^{2}],$\ in the body of the scattering amplitude.

In comparing the solution here attained with the scalar potential valid for a
purely timelike background, given in Ref. \cite{Manojr3}, it is instructive to
point out that both possess a similar logarithmic behavior for $r\rightarrow0$
and $r\rightarrow\infty$. The difference lies mainly in the directional
dependence on the $\beta-$angle, responsible for the anisotropy, absent in the
purely timelike case.

\subsection{The gauge potential}

Although the propagator of the gauge sector is composed by eleven terms, only
six of them will contribute to the scattering amplitude, namely $\theta
^{\mu\nu},S^{\mu\nu},\Lambda^{\mu\nu},T^{\mu}T^{\nu},Q^{\mu\nu},Q^{\nu\mu}.$
This is a consequence of the current-conservation law ($k_{\mu}J^{\mu}=0).$
The first two terms yield, in the non-relativistic limit, the
Maxwell-Chern-Simons (MCS)\ scattering amplitude, already carried out in Refs.
\cite{MCS}. The other four terms lead to background depending scattering
amplitudes. In order to obtain the total scattering amplitude mediated by the
gauge field, one must previously evaluate the following current-current
amplitude terms,
\begin{align}
j^{\mu}(p_{1})(S_{\mu\nu})j^{\nu}(p_{2})  &  =j^{\left(  0\right)  }%
(p_{1})S_{0i}j^{\left(  i\right)  }(p_{2})+j^{\left(  i\right)  }(p_{1}%
)S_{i0}j^{\left(  0\right)  }(p_{2}),\\
j^{\mu}(p_{1})(T_{\mu}T_{\nu})j^{\nu}(p_{2})  &  =j^{\left(  0\right)  }%
(p_{1}) \left[  (\overrightarrow{\text{v}}\cdot\overrightarrow{\text{v}%
})(\overrightarrow{k}\cdot\overrightarrow{k})-(\overrightarrow{\text{v}}%
\cdot\overrightarrow{k})^{2}\right]  j^{\left(  0\right)  }(p_{2}),\\
j^{\mu}(p_{1})\text{ }(\Lambda_{\mu v})j^{\nu}(p_{2})  &  =j^{\left(
i\right)  }(p_{1})[v_{i}v_{j}]j^{\left(  j\right)  }(p_{2}),\\
j^{\mu}(p_{1})\text{ }(Q_{\mu\nu}-Q_{\nu\mu})j^{\nu}(p_{2})  &  =[j^{\left(
i\right)  }(p_{1})v_{i}j^{0}(p_{2})-j^{\left(  l\right)  }(p_{2})v_{l}%
j^{0}(p_{1})](\overrightarrow{\text{v}}\times\overrightarrow{k}), \label{A3}%
\end{align}
which carried out in the non-relativistic limit, with $v^{\mu}=(0,0,$v$)$ and
$k^{\mu}=(0,\mathbf{k),}$ lead to:%

\begin{align*}
j^{\mu}(p_{1})(S_{\mu\nu})j^{\nu}(p_{2})  &  =\mathbf{k}^{2}%
/m-(2i/m)\mathbf{k}\times\mathbf{p},\text{ \ }j^{\mu}(p_{1})(T_{\mu}T_{\nu
})j^{\nu}(p_{2})=\left[  \text{v}^{2}\mathbf{k}^{2}\sin^{2}\alpha\right]  ,\\
\text{ }j^{\mu}(p_{1})(\Lambda_{\mu v})j^{\nu}(p_{2})  &  =-\frac{\text{v}%
^{2}\mathbf{k}^{2}}{4m^{2}}e^{i\theta},\text{ \ }j^{\mu}(p_{1})\text{ }%
(Q_{\mu\nu}-Q_{\nu\mu})j^{\nu}(p_{2})=\frac{\text{v}^{2}\mathbf{k}^{2}}%
{4m^{2}}[1-e^{i\theta}],
\end{align*}
where\textbf{\ }the vector $\mathbf{p}=\frac{1}{2}(\mathbf{p}_{1}%
-\mathbf{p}_{2})$ is defined in terms of the 2-momenta $\mathbf{p}%
_{1},\mathbf{p}_{2}$ of the incoming electrons, and $\theta$ is the scattering
angle in the CM\ frame. The total scattering amplitude associated with the
gauge sector is obviously given by:
\[
\mathcal{M}_{gauge}=\mathcal{M}_{MCS}+\text{ }\mathcal{M}_{\Lambda
}+\mathcal{M}_{TT}+\mathcal{M}_{QQ},
\]
where $\mathcal{M}_{MCS}$ is the Maxwell-Chern-Simons scattering amplitude
(for which contribute the terms $\theta^{\mu\nu},S^{\mu\nu}$of the gauge
propagator) and the other three are explicitly depending on the background, as
exhibited below:%

\begin{align*}
\mathcal{M}_{MCS}  &  =e_{3}^{2}\left\{  \left(  1-\frac{s}{m}\right)
\frac{1}{\mathbf{k}^{2}+s^{2}}-\frac{2s}{m}\frac{i\mathbf{k}\times\mathbf{p}%
}{\mathbf{k}^{2}(\mathbf{k}^{2}+s^{2})}\right\}  ,\text{ }\mathcal{M}%
_{\Lambda}=\frac{e_{3}^{2}s^{2}\text{v}^{2}}{4m^{2}}\frac{\mathbf{k}^{2}%
}{[\mathbf{k}^{2}+s^{2}]\boxtimes(k)}e^{i\theta},\\
\mathcal{M}_{TT}  &  =e_{3}^{2}\text{v}^{2}\frac{\mathbf{k}^{2}}%
{[\mathbf{k}^{2}+s^{2}]\boxtimes(k)}\sin^{2}\alpha,\text{ }\mathcal{M}%
_{QQ}=-\frac{e_{3}^{2}s\text{v}^{2}}{2m}\frac{\mathbf{k}^{2}}{[\mathbf{k}%
^{2}+s^{2}]\boxtimes(k)}[1-e^{i\theta}],
\end{align*}
with the term, $\boxtimes(k)=\left[  \mathbf{k}^{2}(\mathbf{k}^{2}%
+s^{2}+\text{v}^{2}\sin^{2}\alpha)\right]  ,$ being given in Ref.
\cite{Manojr1}. The amplitude $\mathcal{M}_{MCS}$ leads to the well-know
Maxwell-Chern-Simons potential (see Refs.\cite{MCS}),
\begin{equation}
V_{MCS}(r)=\frac{e_{3}^{2}}{\left(  2\pi\right)  }\left[  \left(  1-\frac
{s}{m}\right)  K_{0}(sr)-\frac{2}{ms}[1-srK_{1}(sr)]\frac{l}{r^{2}}\right]  .
\label{VMCS}%
\end{equation}
which presents a purely logarithmic behavior near the origin, namely:
\begin{equation}
V_{MCS}(r)\rightarrow-\left(  e^{2}/2\pi\right)  [1-s/m-sl/m]\ln r,
\label{VMCS2}%
\end{equation}
and a typical $-1/r^{2}$ behavior in the asymptotic limit. This preliminary
MCS\ result will be corrected by the other background depending contributions,
still to be evaluated. Hence, the remaining task consists in carrying out the
Fourier transforms of the three amplitudes above. Starting from the
$\mathcal{M}_{\Lambda}$-amplitude, the corresponding potential is written as follows:%

\[
V_{\Lambda}(r)=\frac{1}{\left(  2\pi\right)  ^{2}}\frac{e_{3}^{2}s^{2}%
\text{v}^{2}}{4m^{2}}\int_{0}^{\infty}\int_{0}^{2\pi}\frac{e^{ikr\cos\varphi}%
}{[\mathbf{k}^{2}+s^{2}][\mathbf{k}^{2}+s^{2}+\text{v}^{2}\sin^{2}\alpha
]}e^{i\theta}kdkd\varphi,
\]
Again, this integral can not be exactly solved, so that the expansion in first
order in v$^{2}/s^{2}$,
\begin{equation}
\frac{1}{[\mathbf{k}^{2}+s^{2}][\mathbf{k}^{2}+s^{2}+\text{v}^{2}\sin
^{2}\alpha]}\simeq\frac{1}{[\mathbf{k}^{2}+s^{2}]^{2}}-\frac{\text{v}^{2}%
\sin^{2}\alpha}{[\mathbf{k}^{2}+s^{2}]^{3}}, \label{App2}%
\end{equation}
must be adopted. Besides this approximation, an important point concerns the
relation existing between the scattering angle $\left(  \theta\right)  $ and
the integration angle $\left(  \varphi\right)  $: $\theta=\left(  2\varphi
-\pi\right)  ,$ which is decisive for the solution of the relevant angular
integration, now read as
\begin{equation}
\int_{0}^{2\pi}e^{ikr\cos\varphi}e^{i\theta}d\varphi=-\left(  2\pi\right)
\left[  J_{2}(kr)\right]  . \label{Angular2}%
\end{equation}
Considering it and stressing that only the first term (on the right hand side)
of Eq. (\ref{App2}) will provide a first order contribution (in v$^{2}$), the
following potential expression comes out:%

\begin{equation}
V_{\Lambda}(r)=\frac{e_{3}^{2}}{\left(  2\pi\right)  }\frac{\text{v}^{2}%
}{4m^{2}}\left\{  -\frac{2}{s^{2}r^{2}}+K_{0}(sr)+\left(  \frac{2}{sr}%
+\frac{sr}{2}\right)  K_{1}(sr)\right\}  , \label{VLamb}%
\end{equation}
whence one notes that in first order the directional dependence on the angle
$\beta$ does not appear in this potential, which exhibits a behavior as $-\ln
r$ (and as $-1/r^{2})$ near (and far from) the origin.

As for the interaction potential related to the $\mathcal{M}_{TT}-$amplitude,
\[
V_{TT}(r)=\frac{e_{3}^{2}\text{v}^{2}}{\left(  2\pi\right)  ^{2}}\int
_{0}^{\infty}\frac{e^{i\overrightarrow{k}.\overrightarrow{r}}\sin^{2}\alpha
}{[\mathbf{k}^{2}+s^{2}][\mathbf{k}^{2}+s^{2}+\text{v}^{2}\sin^{2}\alpha
]}d^{2}\overrightarrow{k},
\]
the integral can not be exactly solved as well, in such a way the expansion
(at first order in v$^{2}/s^{2}),$
\begin{equation}
\frac{\sin^{2}\alpha}{[\mathbf{k}^{2}+s^{2}]\left[  \mathbf{k}^{2}%
+s^{2}+\text{v}^{2}\sin^{2}\alpha\right]  }\simeq\frac{\text{v}^{2}\sin
^{2}\alpha}{\left[  \mathbf{k}^{2}+s^{2}\right]  ^{2}}, \label{App3}%
\end{equation}
must be properly considered. The associated angular integration is given by
Eq. (\ref{Angular1}), so that the resulting potential takes then the form (at
first order in v$^{2})$:%

\begin{equation}
V_{TT}(r)\simeq\frac{e_{3}^{2}\text{v}^{2}}{\left(  2\pi\right)  }\left\{
\frac{c_{1}}{2s^{2}}K_{0}(sr)-\frac{c_{1}}{s^{4}r^{2}}+\frac{\sin^{2}\beta
}{2s}rK_{1}(sr)+\frac{c_{1}}{s^{3}r}K_{1}(sr)]\right\}  , \label{VTT2}%
\end{equation}
where: $c_{1}=-(\cos2\beta).$

One can now solve the last Fourier transformation for the scattering amplitude
$\mathcal{M}_{QQ},$ written as follows:
\begin{equation}
V_{QQ}(r)=\frac{1}{\left(  2\pi\right)  ^{2}}\frac{e_{3}^{2}\text{v}^{2}s}%
{2m}\int_{0}^{\infty}\frac{e^{i\overrightarrow{k}.\overrightarrow{r}%
}(1-e^{i\theta})}{[\mathbf{k}^{2}+s^{2}][\mathbf{k}^{2}+s^{2}+\text{v}^{2}%
\sin^{2}\alpha]}d^{2}\overrightarrow{k}, \label{VQQ}%
\end{equation}
which must be rewritten according to the approximation (\ref{App2}) and solved
making use of the angular integration (\ref{Angular2}), so that one achieves
at first order:
\begin{equation}
V_{QQ}(r)\simeq\frac{e_{3}^{2}}{\left(  2\pi\right)  }\frac{\text{v}^{2}}%
{2m}\biggl\{-\frac{2}{s^{3}r^{2}}+\frac{3}{2}rK_{1}(sr)+\frac{1}{s}%
[K_{0}(sr)+\frac{2}{sr}K_{1}(sr)]\biggr\}. \label{VQQ2}%
\end{equation}
It interesting to point out that the three potentials, $V_{\Lambda}%
,V_{TT},V_{QQ},$ behave at the same way both near and away from the origin.
Indeed, it is easy to show that these potentials behave as a constant for
$r\rightarrow0$, and as $-1/r^{2}$ for $r\rightarrow\infty$. Regarding that
rest mass of the electron represents a large energy threshold before
low-energy excitations usually observed in condensed matter physics, one
should adopt the following condition $m^{2}>>s^{2}.$ Thereby, the potential
$V_{TT}$ turns out proportionally more significant that $V_{QQ}$ and
$V_{\Lambda},$ which is the least relevant one, in accordance the order of
magnitude of the multiplicative factors $\left(  \text{v}^{2}/4m^{2}%
,\text{v}^{2}/2m,\text{v}^{2}\right)  $ which appear in Eqs. (\ref{VLamb}%
,\ref{VQQ2},\ref{VTT2}).

The total gauge potential, $V_{gauge}(r)=V_{MCS}+V_{\Lambda}+V_{TT}+V_{QQ},$
is then written as a complex combination of Bessel functions and $1/r^{2}$
terms, explicitly as:
\begin{align}
V_{gauge}(r)  &  =\frac{e_{3}^{2}}{\left(  2\pi\right)  }\biggl\{\left[
1-\frac{s}{m}+\text{v}^{2}\left(  \frac{1}{2ms}+\frac{1}{4m^{2}}-\frac
{\cos2\beta}{2s^{2}}\right)  \right]  K_{0}(sr)-\left[  \frac{2l}{ms}%
+\text{v}^{2}\left(  \frac{1}{ms^{3}}+\frac{1}{2s^{2}m^{2}}-\frac{\cos2\beta
}{s^{4}}\right)  \right]  \frac{1}{r^{2}}\nonumber\\
&  +\left[  \left[  \frac{2l}{m}+\text{v}^{2}\left(  \frac{1}{s^{2}m}+\frac
{1}{2m^{2}s}-\frac{\cos2\beta}{s^{3}}\right)  \right]  \frac{1}{r}%
+\text{v}^{2}\left(  \frac{s}{8m^{2}}+\text{ }\frac{\text{sin}^{2}\beta}%
{2s}+\frac{3}{4m}\right)  r\right]  K_{1}(sr)\biggr\}. \label{Vgauge}%
\end{align}

Near the origin, this gauge potential is reduced to a simple expression,
\begin{equation}
V_{gauge}(r)\rightarrow-\frac{e_{3}^{2}}{\left(  2\pi\right)  }\left[
1-\frac{s}{m}-\frac{sl}{m}\right]  \ln r, \label{Vgauge2}%
\end{equation}
which corresponds exactly to the limit of the MCS gauge potential, already
established in Eq. (\ref{VMCS2}). This is an expected result, once all the
potentials $V_{\Lambda},V_{TT},V_{QQ}$ behave as a constant in the limit
$r\rightarrow0$. It is still interesting to observe that the gauge potential
derived in the case of a purely timelike background (see Ref. \cite{Manojr1})
also presents this exact dependence, which shows that all background induced
corrections are negligible in the proximity of the origin for both time and
spacelike backgrounds. Far from the origin the Bessel functions decay
exponentially, so that the gauge potential is ruled by the $1/r^{2}$ terms,
which remain as dominant. So, one has:
\[
V_{gauge}(r)\rightarrow-\frac{e_{3}^{2}}{\left(  2\pi\right)  }\left[
\frac{2l}{ms}+\text{v}^{2}\left(  \frac{1}{2m^{2}s^{2}}-\frac{1}{ms^{3}}%
-\frac{\cos2\beta}{s^{4}}\right)  \right]  \frac{1}{r^{2}},
\]
This is also similar to the asymptotic behavior of the pure MCS\ potential,
$-\left(  2l/ms\right)  r^{-2},$ supplemented by background corrections, which
however do not modify the $1/r^{2}$ physical behavior. Such analysis indicates
that the gauge potential is always attractive in the limit $r\rightarrow
\infty,$ since one relies on the approximation $s^{2}>>$v$^{2}$. The
attractiveness of this potential near the origin depends on the sign of the
coefficient $\left(  1-s/m-sl/m\right)  $ in much the same way as it occurs
with the pure MCS potential: it will be attractive for $s>m/(1+l)$ or
repulsive for $s<m/(1+l)$. \ As it is reasonable to suppose that $m>>s,$ once
the rest mass of the electron represents a large energy threshold before the
low-energy excitations usually observed in condensed matter systems, it should
hold the condition $s<m/(1+l),$ compatible with a repulsive gauge potential.
Since the potential is always attractive far from the origin, there must exist
a region in which the potential is negative (a well region) even in the case
$s<m/(1+l).$\ This general behavior is attested in Fig. 1.

The graphics below exhibits a simultaneous plot for the gauge potential
expression and for the pure MCS potential, given respectively by Eqs.
(\ref{VMCS}) and (\ref{Vgauge}):%

\begin{figure}
[h]
\begin{center}
\fbox{\includegraphics[
trim=0.000000in 0.000000in 1.115897in 0.225013in,
height=2.8106in,
width=2.911in
]%
{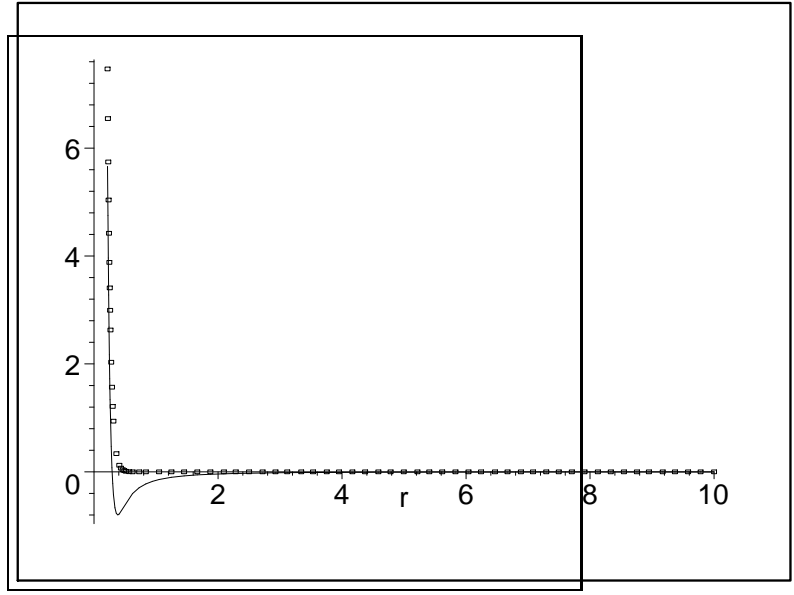}%
}\caption{Plot of the pure MCS potential (box dotted line) $\times$ plot of
the gauge potential (continuos line)
\ \ \ \ \ \ \ \ \ \ \ \ \ \ \ \ \ \ \ \ \ \ for the following parameter
values: $s=20,$ $\mathbf{v}=5,\beta=\pi/2,m_{e}=5.10^{5}.$}%
\end{center}
\end{figure}

Such illustration confirms the equality inherent to the behavior near and away
the origin, at the same time it demonstrates that the presence of the
background may turn this potential attractive at some region. However, this
result is not definitive once it is known that one should address with care
the low-energy potential in order to avoid a misleading interpretation. As
discussed in literature (see Hagen and Dobroliubov \cite{MCS}), in concerning
a nonperturbative calculation one must consider not only the centrifugal
barrier term $\left(  l^{2}/mr^{2}\right)  $, but also the gauge invariant
$\mathbf{A}^{2}-$term coming from the Pauli equation, $\left[
(\overrightarrow{p}-e\overrightarrow{A})^{2}/m_{e}+e\phi(r)-\frac
{\overrightarrow{\sigma}.\overrightarrow{B}}{m_{e}}\right]  \Psi(r,\phi
)=E\Psi(r,\phi).$ The centrifugal barrier term is generated by the action of
the Laplacian operator, $\left[  \frac{\partial^{2}}{\partial r^{2}}+\frac
{1}{r}\frac{\partial}{\partial r}+\frac{1}{r^{2}}\frac{\partial^{2}}%
{\partial\phi^{2}}\right]  ,$ on the total wavefunction $\Psi(r,\phi
)=R_{nl}(r)e^{i\phi l}$; on the other hand, the $\mathbf{A}^{2}$-term is
essential to ensure the gauge invariance in the nonrelativistic domain. As
this term does not appear\thinspace in the context of a nonperturbative
low-energy evaluation, for the same is associated with two-photon exchange
processes (see Hagen and Dobroliubov \cite{MCS}), it must be suitably added up
the low-energy potential in order to assure the gauge invariance. In the
presence of these two terms, the pure MCS\ potential reveals to be really
repulsive instead of attractive. Hence, to correctly analyze the low-energy
behavior of the gauge potential, it is necessary to add up the centrifugal
barrier and the $\mathbf{A}\cdot\mathbf{A}$ term to the gauge potential
previously obtained, leading to the following effective potential:\
\begin{equation}
V_{eff}(r)=V_{gauge}(r)+\frac{l^{2}}{m_{e}r^{2}}+\left(  \frac{e^{2}}{m_{e}%
}\right)  \overrightarrow{A}\cdot\overrightarrow{A} \label{Veff}%
\end{equation}
In order to proceed with this analysis, it is necessary to know the expression
for the vector potential, which was not determined in Ref. \cite{Manojr2}.
This potential may be obtained solving a system of two coupled differential
equations read off from Ref. \cite{Manojr2}, namely: $\nabla^{2}(\nabla
^{2}-s^{2})\overrightarrow{A}=s\overrightarrow{\nabla}^{\ast}\rho
-s[\nabla(\overrightarrow{\text{v}}\times\nabla\varphi)]^{\ast},$ $\nabla
^{2}\varphi+\left(  1/s\right)  (\overrightarrow{\text{v}}\times
\overrightarrow{\nabla})(\overrightarrow{\nabla}\times\overrightarrow{A})=0.$
We proceed decoupling them, yielding the following equation for the vector
potential: $[\nabla^{2}(\nabla^{2}-s^{2})-(\overrightarrow{\text{v}}^{\ast
}\cdot\overrightarrow{\nabla})(\overrightarrow{\text{v}}^{\ast}\cdot
\overrightarrow{\nabla})]\overrightarrow{A}=s\overrightarrow{\nabla}^{\ast
}\rho.$ The solution for this equation (by the usual methods) leads to an
approximated expression in first order in v$^{2}/s^{2}$:%
\begin{align*}
\overrightarrow{A}(r)  &  =\frac{e}{(2\pi)}\biggl\{-\frac{1}{sr}%
(1-\text{v}^{2}/s^{2}\sin^{2}\beta-\text{v}^{2}\cos2\beta/2s^{2}%
)+(1-\text{v}^{2}/s^{2}\sin^{2}\beta+\text{v}^{2}\cos2\beta/2s^{2})K_{1}(sr)\\
&  +\frac{2\text{v}^{2}\cos2\beta}{s^{3}r}K_{0}(sr)-\frac{4\text{v}^{2}%
\cos2\beta}{s^{5}r^{3}}(1-rK_{1}(sr))-\frac{\text{v}^{2}\sin^{2}\beta}%
{2s}rK_{0}(sr)\biggr\}\overset{\wedge}{r}^{\ast}.
\end{align*}
One should now compare the gauge potential (\ref{Vgauge}) with the effective
potential, given by Eq. (\ref{Veff}). In this way, one performs a graphical
analysis of these two functions for small and large electron mass, as it is
shown below:
\begin{figure}
[h]
\begin{center}
\fbox{\includegraphics[
trim=0.000000in 0.099640in 0.217445in 0.225013in,
height=2.7103in,
width=3.813in
]%
{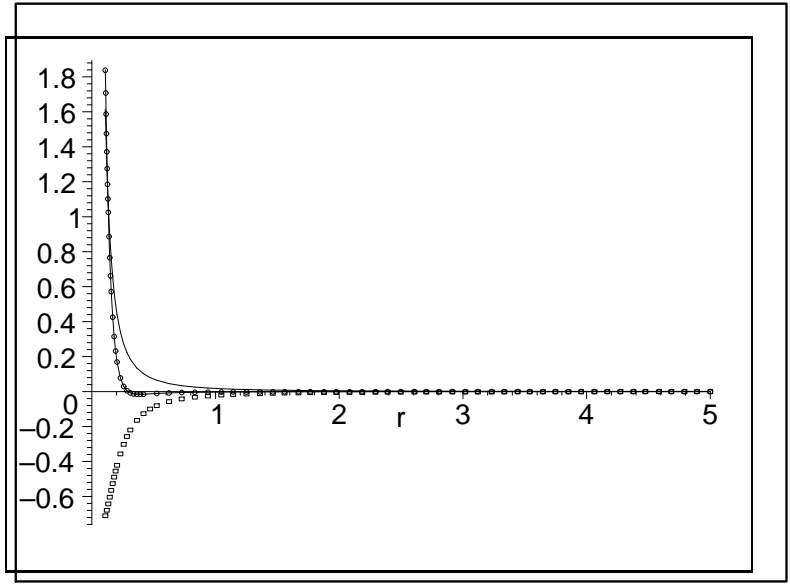}%
}\caption{Plot of the gauge potential (dotted line) $\times$ effective
potential (continuos line) for two set of parameters with distinct mass value:
$\left(  s=20,v=5,m=50,\beta=\pi/2,L=1\right)  $ and $\left(
s=20,v=5,m=5.10^{5},\beta=\pi/2,L=1\right)  .$}%
\end{center}
\end{figure}

For a large mass value $\left(  m_{e}=5.10^{5}\right)  $, one observes that
the effective potential does not differ from the gauge potential (circle
dotted curve), so that both graphics result perfectly overlapped. This fact
reveals that the terms $l^{2}/m_{e}r^{2},$ $A^{2}/m_{e\text{ }}$are not
decisive to alter the behavior of \ the gauge potential in the regime of large
mass $\left(  m_{e}/s\simeq10^{5}\right)  $, On the other hand, for a small
mass parameter $\left(  m_{e}/s\simeq1\right)  ,$ one notes that the gauge
potential (box dotted curve) may differ drastically from the effective
potential (continuos solitary curve). Therefore, in the regime of small mass
the low-energy potential has to be replaced by the effective one in order to
yield the gauge invariant correct behavior, requirement not necessary in the
regime of large mass.

Another point that deserves to be analyzed concerns the influence of the
background direction on the solutions. The graphics in Fig. 3 presents three
simultaneous plots of the gauge potential for different values of the angle
$\beta$:%

\begin{figure}
[h]
\begin{center}
\fbox{\includegraphics[
trim=0.000000in 0.000000in 0.516530in 0.125074in,
height=2.911in,
width=3.5129in
]%
{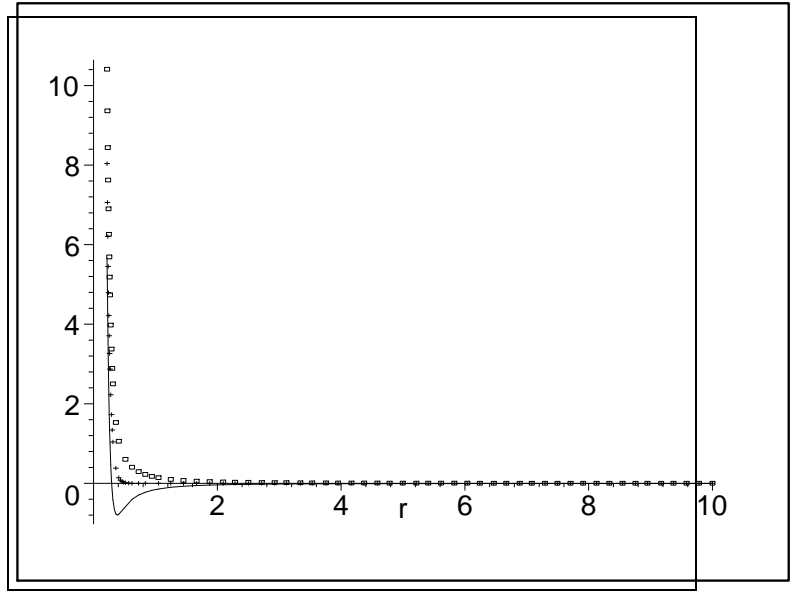}%
}\caption{Plot of the gauge potential for $s=20,$v$=5,m=5.10^{5}$ and
$\beta=\pi/2$ (continuos line$),$ $\beta=3\pi/4$ (box dotted line), $\beta
=\pi$ (cross dotted line). }%
\end{center}
\end{figure}

Such an illustration reflects the anisotropy of the system: depending on the
value of $\beta,$ the potential may become totally repulsive, or exhibit a
region in which is attractive. The interest in such an effect is related to
its possible connection with the anisotropic parameter of order of
high-T$_{c}$ superconductors. An interaction potential whose intensity varies
with a fixed direction indeed leads to an anisotropic wavefunction, which
certainly requires additional investigation.

As a final comment, one should remark that the real potential corresponding
to\ the total $e^{-}e^{-}$ interaction comprises the gauge and the scalar
contributions: $V_{total}(r)=V_{scalar}+V_{gauge}.$ The character attractive
or repulsive of this total potential arises from the combination of these two
expressions for each radial region. Near the origin, for instance, the total
interaction goes as:\ %

\[
V_{total}(r)\rightarrow\frac{1}{\left(  2\pi\right)  }\left\{  -e_{3}%
^{2}\left[  1-\frac{s}{m}-\frac{sl}{m}\right]  +y^{2}\left[  1-\frac
{\text{v}^{2}}{2s^{2}}(1+\sin^{2}\beta)\right]  \right\}  \ln r.
\]
In the regime of \ large mass, the total interaction will be attractive near
the origin whenever the phonic constant $y^{2}$ overcomes the 2-dimensional
U(1) coupling, $e_{3}^{2}$ (or repulsive for $y^{2}<e_{3}^{2}).$ \ Far from
the origin, the total potential exhibits the same logarithmic behavior stated
in Eq. (\ref{Asym}). It should be noted that this asymptotic behavior will
change solely in the case in which a new mass parameter is introduced in, as
it occurs when a spontaneous symmetry breaking takes place.\textbf{\ }This is
mentioned with more details in the Final Remarks. By adjusting the value of
the phonic constant, $y$, one can certainly conclude that the total potential
may always be negative at some region regardless the character of the gauge
interaction, which is a relevant result concerning the possibility of
obtaining $e^{-}e^{-}$ bound states in the framework of this particular model.

\section{Final Remarks}

In this work, one has considered the M\"{o}ller scattering in a planar
Maxwell-Chern-Simons electrodynamics incorporating a Lorentz-violating purely
spacelike background. The interaction potential was calculated as the Fourier
transform of the scattering amplitude (Born approximation) carried out in the
non-relativistic limit. The interaction potential exhibits two distinct
contributions: the scalar (stemming form the Yukawa exchange) and the gauge
one (mediated by the MCS-Proca gauge field). The scalar Yukawa interaction
turns out to be logarithmically attractive and repulsive near and far from the
origin, respectively, in much the same way as verified in the purely timelike
case. As for the gauge interaction, it is composed by a pure MCS potential
corrected by background-depending contributions, which are able to induce
physical interesting modifications despite the smallness of the background
before the topological mass (v$^{2}/s^{2}<<1)$. Near and far from the origin,
this gauge potential goes like the pure MCS counterpart, so that the observed
alterations appear in the intermediary radial region. Namely, it is verified
gauge potential becomes attractive for some values of the parameters. Such
attractiveness remains even in the presence of the centrifugal barrier and
gauge invariant $A\cdot A$ term. Besides the possibility of having a gauge
interaction attractive, it should be mentioned that the total interaction
(scalar plus gauge potential) may always result attractive provided a fine
adjust of the coupling constants values ($y,e_{3})$ is realized. This fact
indeed constitutes a promising result in connection with the possibility of
obtaining the formation of Cooper pairs. It was also reported that in the
regime of a large mass $\left(  m_{e}/s\simeq10^{5}\right)  ,$ the effective
low-energy gauge invariant potential becomes equal to the gauge potential with
a high precision, whereas in the regime of small mas $\left(  m_{e}%
/s\simeq1\right)  $ these two expressions become sensibly different.

The real possibility of obtaining Cooper pairs may be checked by means of a
quantum-mechanical numerical analysis of the non-relativistic interaction
potential here derived. Such potential should be introduced in the
Schr\"{o}dinger equation, whose numerical solution will provide the
corresponding $e^{-}e^{-}$ binding energies for each set of parameter values
stipulated. One should remark that the values must be chosen in accordance
with the usual scale of low-energy excitations in a condensed matter system.
This analysis may be performed for the potentials obtained both in the case of
a purely timelike and spacelike background, having in mind also the issue of
the anisotropy of the resulting parameter of order.

One must now comment on the validity of the approximation which has been here
adopted. At first sight, the higher order terms (in v$^{2})$\ are always
negligible\ before the first order ones. Indeed, this is true for terms that
decay quickly at large distances. Near the origin, although, it might occur
that a high order term (in v$^{2})$\ come to increase with $r$ more rapidly,
overcoming a first order term,\ fact which really is related to its radial
dependence in\ the limit\ $r\rightarrow0$. \ Such a behavior would be observed
if a second order term\thinspace\ (in v$^{2})$\ had a more pronounced power in
$\left(  1/r\right)  $ than the first order one. According to all evaluations
carried out in second order, this fact was not observed, which confirms the
validity of the approximation adopted as well as the outcomes obtained in this
work.\textbf{\ }

The absence of screening, first observed in Refs. \cite{Manojr2},
\cite{Manojr3}, is here manifest only in the scalar potential expression by
means of the asymptotic logarithmic term, once the gauge sector revealed an
asymptotic behavior much different $\left(  \sim1/r^{2}\right)  .$ Some usual
planar models, in (1+2) dimensions, are known for exhibiting a confining
(logarithmic) potential as representation of the gauge interaction, such
behavior however does not reflect a convenient physical interaction, since it
increases with distance. To represent a physical interaction, it may be
changed to a condensating potential, which may be attained when the model is
properly supplemented by new parameters of mass.\ The consideration of the
Higgs mechanism is a suitable tool able to provide a Proca mass to the gauge
field and to induce an efficient screening of the corresponding field
strengths and solutions, bypassing this difficulty. In a recent work
\cite{Manojr4}, it was accomplished the dimensional reduction of an
Abelian-Higgs Lorentz-violating model endowed with the CFJ term, resulting in
a planar Maxwell-Chern-Simons-Proca electrodynamics coupled to a massive
Klein-Gordon field $\left(  \varphi\right)  ,$ with a scalar Higgs sector.
This resulting Higgs-Lorentz-violating planar model has been analyzed in some
theoretical directions, revealing new possibilities and outcomes. The
classical solutions for field strengths $\left(  \mathbf{E},\mathbf{B}\right)
$ and four-potential $\left(  A^{0},\mathbf{A}\right)  $ were carried out for
a static point-like charge, yielding interesting deviations in relation to the
pure MCS-Proca electrodynamics \cite{Manojr5}. A particular feature of this
model is the presence of totally screened modes (all its physical excitations
are massive). Furthermore, one can show that it provides stable charged vortex
configurations\ able to bring about the associated Aharonov-Casher effect
\cite{Belich}.\ The consideration of the M\"{o}ller scattering in this
framework \cite{Manojr6} will certainly lead to an entirely shielded
interaction potential, with the logarithmic term being replaced by K$_{0}$,
K$_{1}$ functions.

\end{document}